\documentclass{emulateapj}
\usepackage{color}

\newcommand{\myemail}{geraint.lewis@sydney.edu.au}
\newcommand{\pandas}{PAndAS}

\definecolor{orange}{RGB}{255,127,0}
\definecolor{purple}{RGB}{255,0,127}

\newcommand{\dist}[1]{${\rm #1\ kpc}$}
\newcommand{\adist}[1]{${\rm \sim#1\ kpc}$}

\newcommand{\vel}[1]{${\rm #1\ km\ s^{-1}}$}
\newcommand{\avel}[1]{${\rm \sim#1\ km\ s^{-1}}$}

\slugcomment{Draft Version:  \today}

\shorttitle{\pandas\ in the mist}
\shortauthors{Lewis et al.}

\begin{document}


\title{\pandas\ in the mist: The stellar and gaseous mass\\ within the halos of M31 and M33}


\author{Geraint F. Lewis$^{1,2,3,4}$,  
Robert Braun$^5$, 
Alan W. McConnachie$^6$, 
Michael J. Irwin$^4$, 
Rodrigo A. Ibata$^7$, 
Scott C. Chapman$^4$, 
Annette M. N. Ferguson$^8$, 
Nicolas F. Martin$^7$,
Mark Fardal$^9$,
John Dubinski$^{10}$,
Larry Widrow$^{11}$,
A. Dougal Mackey$^{12}$,
Arif Babul$^{13}$
Nial R. Tanvir$^{14}$
\&
Michael Rich$^{15}$
 }
\affil{$^3$Sydney Institute for Astronomy, School of Physics A28, The University of 
Sydney, NSW 2006, Australia \\
$^4$Institute of Astronomy, University of Cambridge, Madingley Road, Cambridge, CB3 0HA, U.K. \\
$^5$CSIRO Astronomy and Space Science, PO Box 76, Epping NSW 1710, Australia \\
$^6$Dominion Astrophysical Observatory, 5071 West Saanich Road, Victoria, B.C., V9E 2E7,Canada \\
$^7$Observatoire de Strasbourg, 11, rue de l'Universite, F-67000 Strasbourg, France  \\
$^8$Institute for Astronomy, University of Edinburgh, Blackford Hill, Edinburgh, EH9 3HJ, U.K. \\
$^9$ University of Massachusetts, Dept. of Astronomy, Amherst, MA 01003-9305, U.S.A. \\
$^{10}$ Department of Astronomy \& Astrophysics, 50 St. George St., University of Toronto, M5S 3H4, Canada\\
$^{11}$ Department of Physics, Queen's University, 99 University Avenue, Kingston, Ontario, K7L 3N6, Canada \\
$^{12}$ Research School of Astronomy \& Astrophysics, Mount Stromlo Observatory, Cotter Road, Weston Creek, ACT 2611, Australia\\
$^{13}$ Dept. of Physics \& Astronomy, University of Victoria , Victoria, BC, V8W 3P6, Canada   \\
$^{14}$ Department of Physics \& Astronomy, University of Leicester, Leicester, LE1 7RH, U.K.\\
$^{15}$ Division of Astronomy, University of California, 8979 Math Sciences, Los Angeles, CA. 90095-1562, U.S.A.\\
}


\altaffiltext{1}{Australian Research Council Future Fellow}
\altaffiltext{2}{Email: {\tt \myemail}}


\begin{abstract}
Large scale surveys of the prominent members of the Local Group have provided compelling 
evidence for the hierarchical formation of massive galaxies, revealing a wealth of substructure 
that is thought to be the debris from ancient and on-going accretion events.
In this paper, we compare  two extant surveys of the M31-M33 subgroup of galaxies; the 
Pan-Andromeda Archaeological Survey (PAndAS) of the stellar structure, and a combination
of observations of the \ion{H}{1} gaseous content, detected at 21cm. 
Our key finding is a marked lack of spatial correlation between these two 
components on all scales, with only a few potential overlaps between stars and gas.
The paucity of spatial correlation significantly restricts the analysis of kinematic correlations, 
although there does appear to the \ion{H}{1} kinematically associated with the Giant Stellar Stream 
where it passes the disk of M31.
These results demonstrate that that different processes must significantly influence the dynamical evolution of the 
stellar and \ion{H}{1} components of substructures, such as ram pressure driving gas away from a purely gravitational path.
Detailed modelling of the offset between the stellar and gaseous substructure will provide a determination
of the properties of the gaseous halo of M31 and M33.
\end{abstract}


\section{Introduction}\label{intro}
In $\Lambda$CDM cosmologies, galactic halos are built up over time through the accretion 
and cannibalisation of smaller mass systems. 
Given the complexity of the dynamical interactions, these are  studied in 
detail using  computer simulations~\citep[e.g.][]{1998ApJ...500..575I,2008ApJ...682L..33F}, 
although, due to the difficulties in  modelling baryonic physics, these typically only consider the
evolution of dark matter. However, these predict that the outer region of galactic halos 
should be 
dominated by extensive tidal streams, whereas the shorter dynamical time scales of the
inner halo result in complete destruction and a smooth stellar distribution
\citep[e.g.][]{2005ApJ...635..931B,2010MNRAS.406..744C,2011MNRAS.416.2802F}.

\begin{figure*}
\begin{center}
\includegraphics[scale=1.44, angle=0]{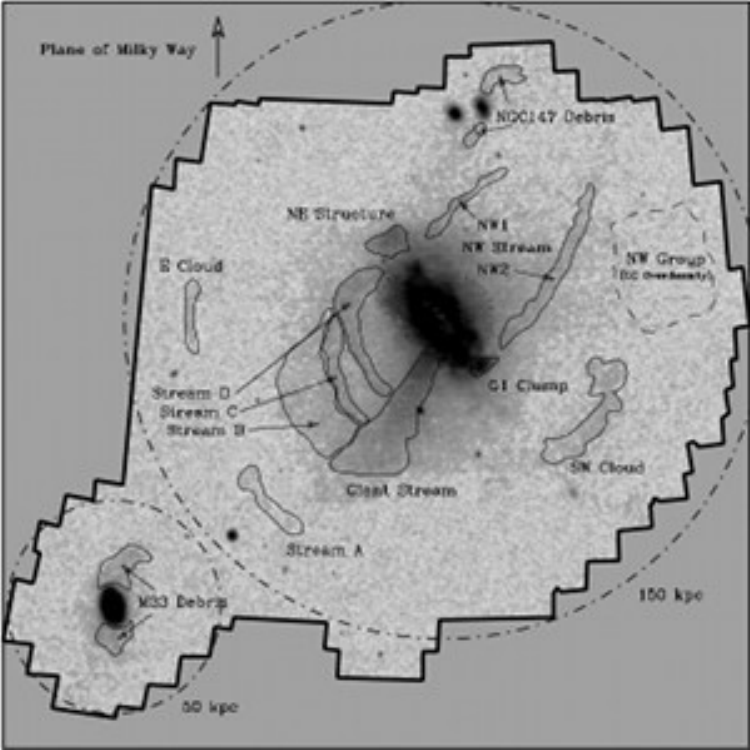}
\caption{A schematic illustration of the prominent substructure apparent in Red Giant Branch (RGB) 
stars; this encompasses a broad cut in metallicity and nonlinearly scaled to enhance the stellar
substructure. 
The thick solid line represents the PAndAS footprint, with the thinner solid lines resenting the main stellar
substructure; as well as structure in the halo of M31, there is also tidal debris associated with M33 and NGC147.
A number of substructures have been presented in previous publications (see Section~\ref{stars_2}), 
but several additional features are apparent in this full map, including 
the Eastern Cloud and the broad swath of stars associated with Stream D \citep[see also the PAndAS map presented in][]{2011ApJ...732...76R}.
The dashed line represents
an over-density of globular clusters identified in \citet{2010ApJ...717L..11M}. Closer in to the disk of M31, 
the dotted lines represents  the inner shells and loops thought to be wraps of the Giant Stellar Stream.
The large dot-dashed circle is at a radius of \dist{150} from the centre of M31, and the smaller
being \dist{50} from the centre of M33.
}\label{Figure1_Whole_Maps}
\end{center}
\end{figure*}

The progenitors of substructure could possess a mix of stars and gas, generally more concentrated than
the dark matter halos in which they reside. The precise composition of any in-falling halo will depend upon
its mass and evolutionary history, and we would expect dwarf galaxies to be stripped of their gas due to
interactions within the Local Group, whereas larger galaxies will hold-on to its gaseous component.
While stars are effectively collisionless, more complex internal physics  influences
the evolution of gas, including internal shocking, cooling and collapse, star formation and its associated
feedback, as well as  ram-pressure stripping due  to the presence of hot galactic halo gas
\citep[e.g. see][]{1994MNRAS.270..209M,1999MNRAS.309..161M,2008MNRAS.390L..24B,2009MNRAS.399.2004M}.
This is quite apparent in the two major accretion events within the Milky Way halo, with the body and 
stream of the Sagittarius Dwarf Galaxy possessing no gas, having potentially lost it during its 
initial interaction with the Galaxy (\citealt{1999A&A...349....7B}, but see also \citealt{2004ApJ...603L..77P}), 
while the Magellanic Stream appears to be completely devoid of stars, consisting of  stripped gas, although the question of whether this is the result of tidal or 
ram pressure forces has yet to be decided \citep{2008ApJ...680..276S,2010ApJ...721L..97B,2011PASA...28..117D}. 
However, while we should expect different distributions for
various disrupted components in M31 and M33, the complexities of ram-pressure stripping, which may
require full magneto-hydrodynamic approaches to simulate \citep{2012Ruszkowski}, makes drawing  
robust conclusions difficult. 

We are now in an era where large scale surveys are providing a global picture 
of the formation and evolution of galaxies over cosmic time \citep[e.g. see][]{2001Sci...293.1273A}. 
Given their
distances, unraveling the fine details of galactic evolution is below the resolution 
achievable with the vast majority of galaxies in the Universe. This is unfortunate, 
as it is on small scales, with the structure within the halos of large galaxies, that
have presented the major challenge to the prevailing ${\rm \Lambda}$CDM 
paradigm \citep[e.g. the missing satellite problem;][]{1999ApJ...522...82K,1999ApJ...524L..19M}.

Luckily, the large members of the Local Group, the Milky Way, the Andromeda (M31)
and Triangulum (M33) galaxies, are close enough for such fine scale detail
to be resolvable, although their immense angular scale presents a significant challenge 
to the observability of their extensive halos. To this end, the last
decade has seen the advent of large, detailed surveys of both the stellar and gaseous 
components of Local Group members,  for the first time providing a panoramic view of galaxy formation 
in action; for the Milky Way, this includes surveys such as RAVE \citep{2011AJ....141..187S},
SEGUE \citep{2010ApJ...714..663D}, GASS \citep{2009ApJS..181..398M} and others.

In this paper, we present a detailed comparison of the stellar and 
gaseous matter in the halos of our nearest large companions in the Local Group, 
M31 and M33, using a new  map of stellar structure obtained through 
the Pan-Andromeda Archaeological Survey (PAndAS) and the most detailed maps of \ion{H}{1}, 
obtained during several telescope campaigns. 
In  Section~\ref{observations}, we outline the observational campaigns that resulted 
in the data presented, with a summary of their key scientific results. In Section~\ref{correlations} we 
compare the spatial and kinematic distributions of gaseous 
and stellar substructure. We present the interpretation of these correlations in 
Section~\ref{interpretation} and our conclusions in Section~\ref{conclusions}.

\begin{figure*}
\begin{center}
\includegraphics[scale=1.7, angle=0]{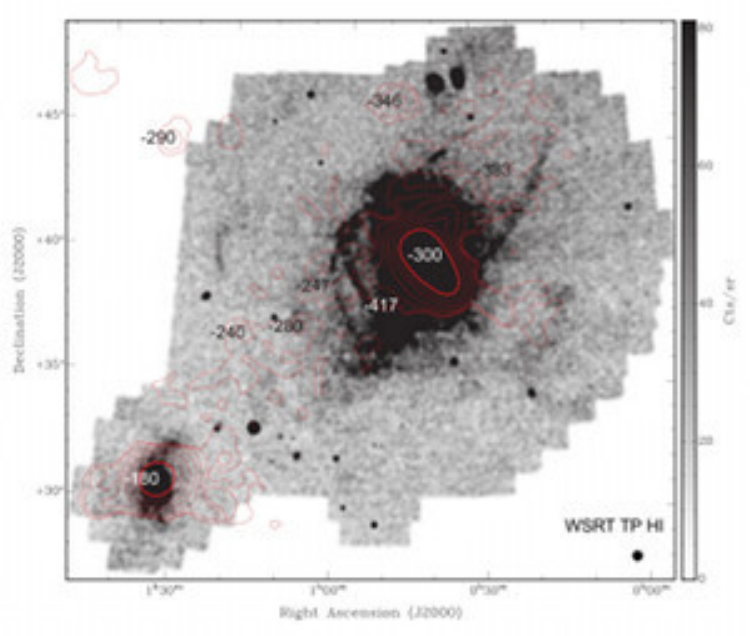}
\caption{The distribution of \ion{H}{1} gas (red contours), drawn from
  the survey of \citet{2004A&A...417..421B}, overlaid on a surface
  density map of stars identified as being on the RGB within M31 and
  M33 in the entire PAndAS footprint  \citep[see][]{2009Natur.461...66M,2011ApJ...732...76R}. The logarithmic contours are drawn at
  integrated column densities of \ion{H}{1}, with N$_{HI}$~=~$10^{17}$ to $10^{20}$
  cm$^{-2}$ in steps of 0.5 dex, and the annotations indicate the heliocentric radial velocity
  of \ion{H}{1} features. The stellar map again represents a broad range of metallicities and has a non-linear stretch to accentuate 
  the substructure. }
\label{Figure2_Whole_Map_HI}
\end{center}
\end{figure*}

\section{Observations \& Properties}\label{observations}
Some galaxies of the Local Group have been known since pre-history 
\citep[e.g. see Section 3.1 of][]{2006MNRAS.366..996G}\footnote{
Given its proximity, M31 is one of the few galaxies visible to the naked eyes. 
Its existence was first documented by the Persian astronomer Abd al-Rahman 
Al-Sufi in his treatise on stellar astronomy titled ÒKitab al-Kawatib al-Thabit al-MusawwarÓ 
(Book on the Constellations of the Fixed Stars), published in AD 964, where he both identified 
its position in the sky and summarized his observations.}, 
and have played a pivotal role in our explorations of the workings
of the Universe. Reproducing a detailed account of this history is beyond the 
scope of this article, and the interested reader is directed to recent reviews 
\citep[e.g.][]{2009ARA&A..47..371T,2010AN....331..526W,2010arXiv1012.2229T,alan2012a}. 
In the following we will focus upon the recent optical and radio surveys of M31
and M33, with a review of the observational programs and key scientific discoveries to date.

\subsection{Stellar Observations}\label{stars}

\subsubsection{Observational Program}\label{stars_1}
The Pan-Andromeda Archaeological Survey (PAndAS) is a survey of 
400 square degrees, covering the halo of M31 out to a distance of \dist{150}, and 
M33 to a corresponding distance of \dist{50}, undertaken as 
a large program on  MegaCam\footnote{\tt cfht.hawaii.edu/Instruments/Imaging/Megacam}, mounted
on the 3.6-m Canada-France-Hawaii Telescope (CFHT). 
Integrations were sufficient to achieve photometric limits 
$g=25.5$ and $i=24.5$ at a $S/N\sim10$, 
reaching several magnitudes below the tip of the Red Giant Branch
(RGB) at the distance of M31/M33 
\citep[D\adist{780-900};][]{2011ApJ...740...69C}\footnote{In 
the following, we adopt the distances $D_{M31} =$ \dist{779} and $D_{M33} = $ \dist{820}, 
each with an uncertainty of $\pm$ \dist{20}, as derived by \citet{2012Conn}.}.
The photometric data-taking was completed in early 2011, 
with the first published map, covering roughly half of the total observed area,  presented in 
\citet{2009Natur.461...66M}, with the (almost) entire dataset  first presented in \citet{2011ApJ...732...76R}.
The final map and high level data products will be made publicly available in a forthcoming publication \citep{alan2012a,alan2012}.

This study focuses upon Red Giant Branch (RGB) stars at the distance of M31 and M33, 
selected with cuts in color and magnitude 
\citep[see][]{2001Natur.412...49I,2007ApJ...671.1591I,2009Natur.461...66M}.
Figure~\ref{Figure1_Whole_Maps} presents the distribution of RGB stars, overlain 
by a schematic map of the  prominent stellar substructure; note that a non-linear scaling has
been applied to the RGB density, to bring out faint substructure, and contains a broad swath 
of metallicities to reveal metal-poor $(-3.0<[Fe/H]<-1.7)$, intermediate $(-1.7<[Fe/H]<-0.7)$ 
and metal-rich $(-0.7<[Fe/H]<0.0)$ substructure \citep[c.f.][]{2007ApJ...671.1591I}.
The thick solid line is the entire PAndAS footprint, with 
stellar substructure as labelled thin solid lines. The dashed curve represents 
a significant over-density of globular clusters identified as the NW Group by \citet{2010ApJ...717L..11M}.
The large dot-dashed line corresponds to a circle of  radius of \dist{150} from the center of M31, whereas the 
smaller dot-dashed circle represents a distance of \dist{50} from the center of M33. 
It is very apparent that accompanying M31 and M33 is a 
wealth of substructure consisting of extensive streams and dwarf galaxies [these will be discussed in more 
detail in Section~\ref{stars_2}, and see \citet{2011ApJ...732...76R}].

The stellar catalog derived from  PAndAS 
is built upon earlier observations with the
CFHT/MegaCam \& CHFT/CFH12k \citep[see][]{2003MNRAS.343.1335M,2007ApJ...671.1591I}. 
In parallel, 
a number of fields, targeting prominent substructure and dwarfs, were targeted with 
DEIMOS \citep{2003SPIE.4841.1657F} on the 10m Keck-II Telescope
\citep[e.g.][]{2008MNRAS.390.1437C,2011MNRAS.417.1170C}.
With a moderate resolution ($R\sim6000$), observations of 60-90\ mins around the 
prominent CaT absorption lines  $\sim8600{\rm \AA}$ resulted in a S/N$\sim5$ 
for targets of $i\sim21$, with a corresponding velocity resolution of \avel{5-10};
these will be described in more detail in Section~\ref{kine}.

 \begin{figure*}
\begin{center}
\includegraphics[scale=1.2, angle=0]{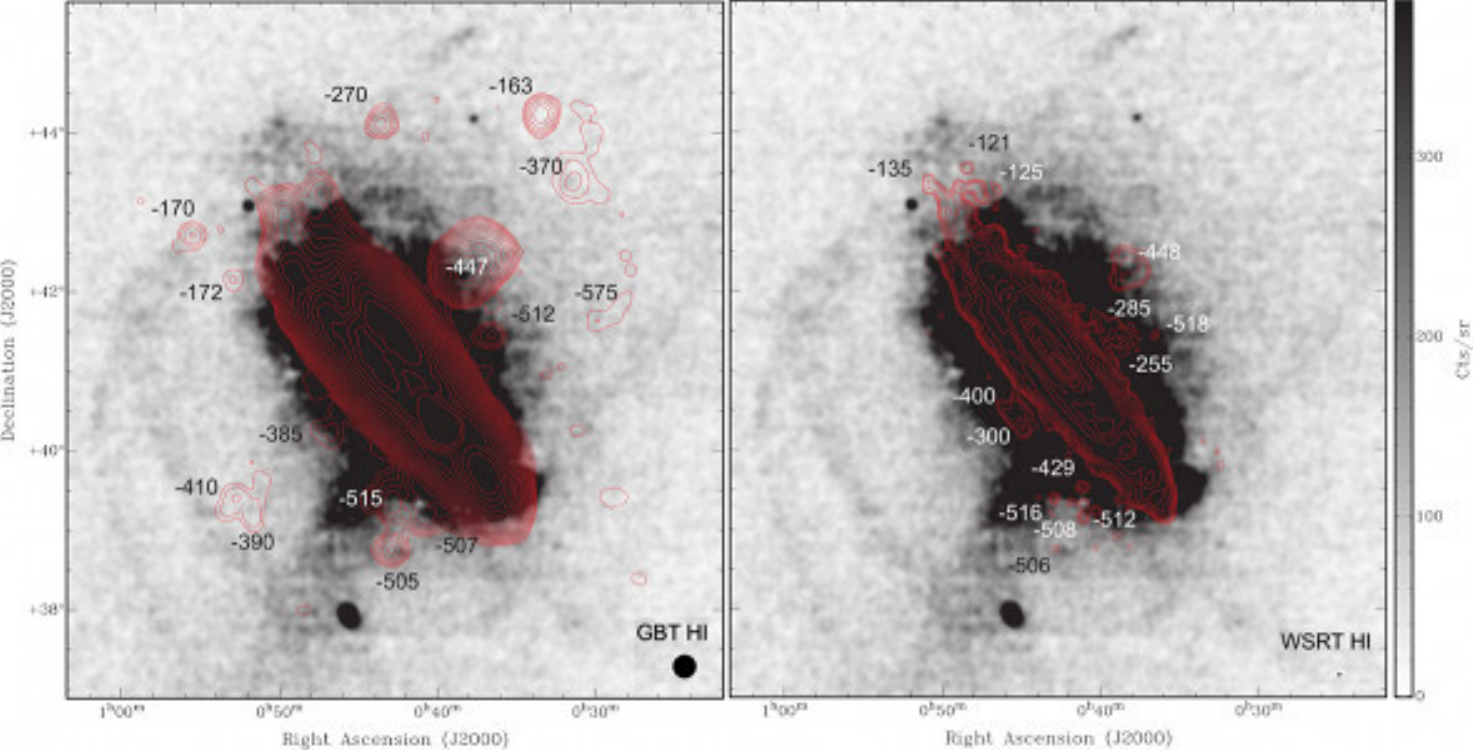}
\caption{As Figure~\ref{Figure2_Whole_Map_HI}, but focused upon the main disk and
  inner halo substructure of M31.  The left-hand panel is overlain
  with \ion{H}{1} contours drawn
  from the study of \citet{2004ApJ...601L..39T} which reveal structure
  at a spatial resolution of 18 arcmin. The
  logatithmic contours are drawn at peak brightness temperatures
  T$_{B}$~=~0.025 to 20~K in steps of 0.15 dex.  
    The right-hand panel is overlain with the 
  \ion{H}{1} observations of \citet{2009ApJ...695..937B}, further revealing smaller scale
  structure, at 4 arcmin resolution, in the gaseous component. The logarithmic contours are drawn at
  integrated column densities N$_{HI}$~=~$7\times10^{18}$ to $4\times10^{21}$
  cm$^{-2}$ in steps of 0.28 dex.  
  The annotations on both panels indicate the heliocentric radial velocity
  of \ion{H}{1} features, and the image is \dist{108} across. } 
  \label{Figure3_M31}
\end{center}
\end{figure*}

\subsubsection{Stellar Properties}\label{stars_2}
Detailed analysis of the stellar substructure in PAndAS will be presented in \citet{Ibata2012}
and \citet{alan2012}, so here we focus on the key features.

There is a wealth of substructure apparent in the outer disk/inner halo regions of M31, 
revealed in the earlier Wide Field Camera 
on the  2.5-m Isaac Newton Telescope survey of the galaxy
\citep[][\& see Figure~\ref{Figure1_Whole_Maps}]{2001Natur.412...49I,2008AJ....135.1998R},
with the 
most significant structure  being an extensive stellar stream, 
the Giant Stellar Stream (GSS), in the halo of M31.
This is seen as a metal-rich substructure wrapped within a more metal-poor halo, and is apparently 
wraps around the disk of M31 in the north-east 
\citep[e.g.][]{2002AJ....124.1452F,2007ApJ...671.1591I,2012MNRAS.423.3134F}.

Accompanying the GSS is a series of streams, labelled $B$, $C$ and
 $D$, which lie perpendicular to the Giant Stream \citep{2007ApJ...671.1591I}, plus Stream A, a stellar
 over-density \adist{125} (in projection) from M31.
 Further features include the South West (SW) Cloud and the North West 
(NW) Stream \citep{2009Natur.461...66M,2010ApJ...717L..11M}.
As revealed in Figure~\ref{Figure1_Whole_Maps}, and as shown by \citet{2011ApJ...732...76R}, 
this latter feature appears to loop back towards 
M31 as the North East (NE) Stream, and intercepts the And\ XXVII dwarf galaxy; 
following \citet[][]{2011ApJ...731..124C}, we identify this entire feature as the North West Stream, and 
label the two components of it as NW1 and NW2.
Finally, also apparent in this image  is another significant over-density of stars, named the Eastern (E) Cloud, 
located a distance of \adist{125} from M31.

As well as the extensive substructure, the PAndAS observations revealed the presence of a 
smooth stellar halo of M31 out to \dist{150} (\citealt{2007ApJ...671.1591I}; see also the SPLASH characterisation
of the smooth halo by \citealt{2012arXiv1210.3362G}).
Furthermore, the
data reveals a wealth of dwarf galaxies  
\citep[e.g.][]{2006MNRAS.371.1983M,2007MNRAS.380..281M,2011ApJ...732...76R}, 
globular clusters and extended clusters  
\citep{2008MNRAS.385.1989H,2011ApJ...730..112C,2011MNRAS.414..770H,2012MNRAS.tmp.2635T}; 
while these extended clusters (ECs) are more diffuse than normal globular 
clusters \citep{2005MNRAS.360.1007H}, they appear to have
very similar stellar populations \citep{2006ApJ...653L.105M,2007ApJ...655L..85M}, 
and do not appear to be dominated by dark matter \citep{2009MNRAS.396.1619C}.

   The structure around M33, apparent as a distorted outer disk, was originally presented by 
 \citet{2009Natur.461...66M,2010ApJ...723.1038M}. 
 This structure is quite clearly asymmetric about M33, being
 significantly more prominent in the north than in the south. 
 
 The extension of the PAndAS 
 survey towards to north of M31, in the region of the two satellite galaxies, NGC147 and NGC185, 
 reveals the presence of stellar debris that appears to have been torn from NGC147, as well
 as a new dwarf galaxy, Cass II (also known as And\ XXX), which is a potential satellite of the 147/185 subgroup, and
 will be discussed in more detail in a forthcoming publication \citep{irwin2012}.

\subsection{\ion{H}{1} Observations}\label{gas}

\subsubsection{Observational Program}\label{hi_1}
The extended environment of M31 and M33 was observed by
\citet{2004A&A...417..421B} using the Westerbork Synthesis Radio
Telescope (WSRT) as fourteen single dish telescopes. A region of
$60\times30^\circ$ in RA$\times$Dec was imaged in the \ion{H}{1}
emission line at an effective resolution of 49 arcmin with an RMS
sensitivity corresponding to a column density of $5\times
10^{16}$cm$^{-2}$ over \avel{30}. This unprecedented sensitivity
permitted detection of \ion{H}{1} in emission from column densities
that have previously only been probed by Ly$\alpha$ absorption toward
background QSOs. This low resolution survey has been supplemented by
higher resolution total power observations made with the Green Bank
Telescope (GBT). A region covering $7^\circ \times7^\circ$ centered on M31
and $5^\circ \times5^\circ$ centered on M33 was observed with GBT during 2002
October with multiple perpendicularly scanned coverages yielding
images with angular resolution as high as 9 arcmin. The M31 data have
been presented previously by \citet{2004ApJ...601L..39T}, while the
M33 data were obtained with the same setup and have been reduced in
a similar fashion. These moderate resolution data are supplemented
with interferometric mosaic observations consisting of a 163 pointing
WSRT coverage of M31 \citep{2009ApJ...695..937B} and Very Large Array
(VLA) coverage of M33 consisting of 6 pointings in the B and CS
configurations and a 20 pointing mosaic in the D configuration. Some
early M33 results based on the B and CS configuration data were
presented in \citet{2002ASPC..276..370T} while an independent
reduction of all three VLA configurations has been presented by
\citet{2010A&A...522A...3G}. The interferometric data provides
resolution as high as 15~arcsec in M31 and as high as 5~arcsec in M33.

\subsubsection{\ion{H}{1} Properties}\label{hi_2}
The \ion{H}{1} distribution in the extended M31 and M33 environment is
overlain as contours on the surface density of RGB stars in the PAndAS
survey in Figure~\ref{Figure2_Whole_Map_HI}. The \ion{H}{1} contours delineate the
diffuse gaseous filament that connects M31 and M33 as well as other
filamentary features extending both to the northwest of the M31 disk
and to the southwest. Heliocentric radial velocities of various
features are indicated with the annotations. The intrinsically diffuse
nature of these features has been verified with follow-up observations
using the GBT directed at the brightest positions along the M31-M33
filament \citep{2004A&A...417..421B}. Despite employing a 25 times
smaller beam area in these GBT observations, the features were
detected at the same low column densities of only a few times
$10^{17}$cm$^{-2}$. This rules out the possibility that the filament
is simply a collection of unresolved clumps in the discovery
observations. Several discrete features are also seen in the northern
portion of the field. It is noteworthy that the M31-M33 filament
connects the systemic heliocentric velocities of M31 (\avel{-300}) and M33
(\avel{-180}).

The more immediate \ion{H}{1} environment of M31 is illustrated in
Figure~\ref{Figure3_M31} where the GBT image (left-hand panel) and WSRT mosaic 
(right-hand panel) are
similarly overlain on the PAndAS survey. At these higher angular
resolutions, of 18 and 4 arcmin respectively, much of the diffuse gas is not detected
but rather only the discrete High Velocity Cloud (HVC) features within about
\dist{100} of the disk. Of note is that the radial velocity of discrete
HVC features follows the basic pattern of disk rotation with the most
negative velocities occurring in the southwest and most positive in the
north-east. 

\begin{figure*}
\begin{center}
\includegraphics[scale=1.1, angle=0]{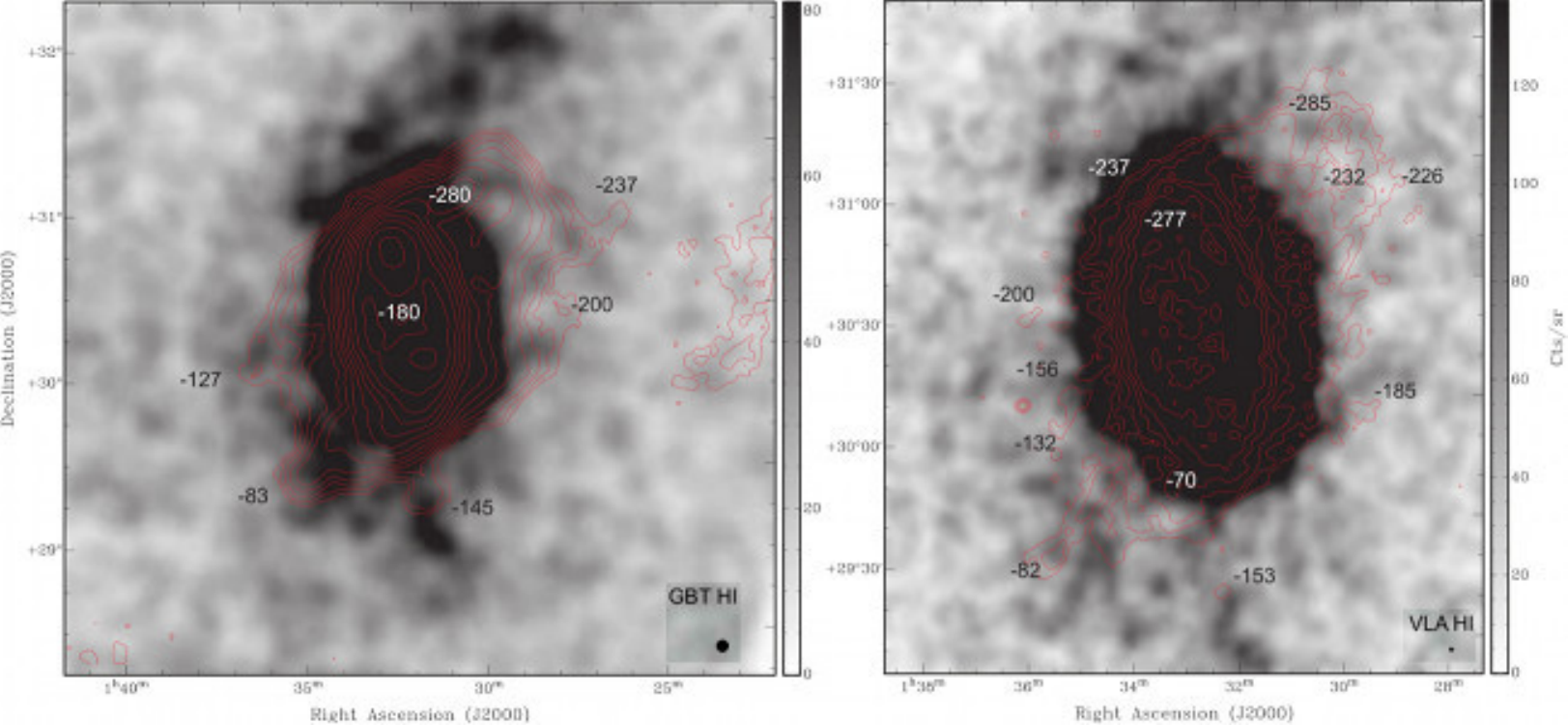}
\caption{The RGB stellar map of M33 and its associated tidal structure, overlain with the 
\ion{H}{1} observations of \citet{2004ApJ...601L..39T} with the GBT, with a resolution of 9 arcmin 
 (left-hand panel), with logarithmic contours are  drawn at peak brightness temperatures 
 T$_{B}$~=~0.15 to 20~K in steps of 0.18 dex. 
The right-hand panel presents the VLA observations of \ion{H}{1} by \citep{2002ASPC..276..370T} 
at 2 arcmin resolution. The logarithmic contours are drawn at peak brightness temperatures
T$_{B}$~=~0.9 to 30~K in steps of 0.26 dex. The annotations indicate the heliocentric radial velocity
of \ion{H}{1} features, and the left-hand image is \dist{60} across, whereas the right-hand image is
\dist{35} across.}
\label{Figure4_M33}
\end{center}
\end{figure*}

The immediate gaseous environment of M33 is illustrated in
Figure~\ref{Figure4_M33} where the GBT image at 9 arcmin
resolution (left-hand panel) and the VLA mosaic with 2 arcmin resolution 
(right-hand panel) are overlain on
the the PAndAS survey. M33 is not as rich in discrete HVC features as
M31 but rather displays a very strong clumpy warp of the outer
\ion{H}{1} disk oriented toward the northwest and southeast, strongly
suggestive of accretion fuelling of the M33 disk by fall-back from the
M31-M33 filament.

\section{Correlations}\label{correlations}

\subsection{Spatial Correlations}\label{space}

\subsubsection{The disk \& halo of M31}\label{halom31_space}
Firstly, we consider the large scale distribution of RGB stars and \ion{H}{1} gas presented in
Figure~\ref{Figure2_Whole_Map_HI}, 
focusing upon material in the vicinity of M31. Leaving aside the inner disk region,
which will be examined in detail shortly, it is useful to consider material along the NW axis, connecting
M31 and M33, and that which lies off this axis in the the halo of M31.

It is clear that  that the eastern portion of the halo possesses  
one substantial stellar substructure, namely the Eastern Cloud, as well as a number of dwarf 
galaxies \citep[see][]{2011ApJ...732...76R}. This section of the halo is effectively devoid of a substantial
quantity of \ion{H}{1} material, except for a lone clump located to the far north, and some lower level material closer
to the disk of M31, with no significant enhancement in the vicinity of any of the dwarf galaxies.
Hence, there appears to be no substantive correlation between the stellar and gaseous material in this 
area of the halo.

The situation is very similar in the western halo, with a one extensive stellar substructure, the SW Cloud, 
and a population of dwarf galaxies, and some potential, but low level, substructure features. Note that 
\citet{2010ApJ...717L..11M} also identified a potential over-density of globular clusters,  the NW Group,
which is not associated with an over-density in the stellar density (see Figure~\ref{Figure1_Whole_Maps}). 
Again, this area is almost devoid of significant \ion{H}{1} detections, other than a large spur at 
$\sim10^{\rm o}$ to the major axis, extending \adist{65} from the center of M31. 
Again, there is apparently little stellar substructure along the length of this gaseous 
spur, except at its  tip where it overlies a portion of the SW Cloud. The physical implications of this 
association will be discussed in Section~\ref{interpretation}.

The most conspicuous stellar and gaseous substructures lie along the NW-axis, along the line joining M31 and 
M33. In the region south of M31, there is significant stellar substructure in the 
form of the GSS, and perpendicular steams, B, C \& D, close to the outer disk of M31, and the distant stream A. 
The gas in this region is significantly more extended, lying all along the axis. There is no 
significant overlap of \ion{H}{1} with the GSS, with the gaseous material extending from the disk of M31
over the inner stream fields and towards M33. 
Significant \ion{H}{1} also extends northwards from the disk of M31, over the region encompassing the NW stream, although
the gaseous material is more extensively spread than the stellar substructure in this area.

The two panels in
Figure~\ref{Figure3_M31} zoom in to  the outer disk region of M31, 
overlain with \ion{H}{1} at the two resolutions discussed earlier. 
While the GSS and Stream C are apparent as stellar substructures, 
with Stream B  stretching back towards M31, there appears to be no significant correlation
between the stellar material, and \ion{H}{1} gas. Intriguingly, there is some \ion{H}{1} emission, apparent as 
population of distinct clouds, aligned
with the Giant Stellar Stream, but offset by \adist{15}; as will be discussed in Section~\ref{halom31_kine},
this emission is also kinematically correlated with the GSS \citep[see][]{2009ApJ...695..937B}.

Clearly, the \ion{H}{1} associated with the stellar disk possesses similar distortions 
to the underlying stars, including a Northern Spur, located in the outer disk of M31 (close to the NE Structure
in Figure~\ref{Figure1_Whole_Maps}, see also Figure 1 in \citet{2008AJ....135.1998R}) 
and a potential over-density near the G1-clump; 
however, it is difficult to discern whether either the stellar or gaseous material has 
recently been accreted or is actually disk 
material that has been tidally distorted due to interactions \citep{2005ApJ...634..287I,2007AJ....133.1275F}.

\subsubsection{The disk \& halo of M33}\label{halom33_space}
At the resolution provided by the \citet[][Figure~\ref{Figure2_Whole_Map_HI}]{2004A&A...417..421B}, 
the \ion{H}{1} emission from M33 is seen to be quite extensive and orientated E-W across 
the galaxy. The stellar substructure, on the other hand, lies mainly on the NW-axis, joining 
the disk of M33 at the northern tip of the galaxy.

The situation becomes clearer in Figure~\ref{Figure4_M33}, which focuses upon the stellar distribution of M33
overlain with \ion{H}{1} emission seen with the GBT (left-hand panel) and VLA (right-hand panel)
\citep[][]{2004ApJ...601L..39T,2002ASPC..276..370T}. The striking 
feature in both the \ion{H}{1} and stellar material is the substantial stream pointing 
towards the NW, as well as a less pronounced component in the SE; this 
extended \ion{H}{1} emission was discussed by \citet{2009ApJ...703.1486P}, with the conclusion 
that it results from the  tidal interaction between M33 and M31.
A similar conclusion  with regards to the stellar component was reached by 
\citet{2009Natur.461...66M}, also presenting a numerical model for the tidal interaction of the two galaxies 
which results in a consistent stellar feature.
However, while the stellar and gaseous material possess a similar alignment, 
they are distinct spatially, with the majority of the gas streaming from the Western edge of M33, 
whereas the stellar material streams from the Northern tip of the galaxy. Furthermore, while 
both the Northern and Southern stellar and gaseous components are 
asymmetrical, it is apparent that the stellar component possesses a much more pronounced 
asymmetry.

\subsection{Kinematic Correlations}\label{kine}
While the radio observations provide a global picture of velocities, the nature of multi-object optical
spectroscopy ensures that the stellar kinematics are determined in a series of discrete fields.
As well as the parallel kinematic survey undertaken by members of the PAndAS team 
\citep[e.g.][]{2004MNRAS.351..117I}, significant effort has been undertaken by other groups
\citep[e.g. the SPLASH survey][]{2009ApJ...705.1043K,2009ApJ...705.1275G,2010ApJ...711..671K}. 
However, the general lack of distinct spatial correlations (see Section~\ref{space}) 
means that direct comparison of kinematics at specific locations is not possible, 
and only a general comparison can be undertaken.

\subsubsection{The disk \& halo of M31}\label{halom31_kine}
The most prominent stellar feature in the halo of M31 is the GSS, lying  
close to the (3-D) distance where it meets the spiral disk,
sweeping backwards to over \adist{100} behind~\citep{2001Natur.412...49I,2003MNRAS.343.1335M,2012MNRAS.423.3134F}.
A stellar kinematic survey of the GSS  
shows a strong velocity gradient \citep{2004MNRAS.351..117I}; 
the most distant tip of the stream, 4.5 degrees away from  
M31's disk, have velocities of $v_h$\footnote{In the following, $v_h$ refers to heliocentric velocities,
whereas $v_{M31}$ is relative to M31}\avel{-300}, placing it essentially at rest with regards to M31,
while those approaching the disk are travelling at $v_h$\avel{-500} ($v_{M31}$\avel{-200}).
While there is no apparent gas associated with the extent of the GSS (Section~\ref{space}), 
there is significant parallel \ion{H}{1} emission. 
This gas possesses a strong
velocity gradient, but different to that of the Giant Stellar Stream 
(Figure~\ref{Figure2_Whole_Map_HI}); close to the M31 disk, the gas is moving
close to systemic velocity of M31, whereas away from the disk, the velocity becomes more positive.
The velocity gradient  bridges the systemic velocities of M31 and M33, demonstrating a 
direct connection and common origin.

North of M31, along the M31-M33 axis, there is significant \ion{H}{1} emission which seemingly overlaps with  
North-Western Stream. The velocity of the \ion{H}{1} ($v_h$\avel{-380}) appears to be a continuation 
of the gaseous stream 
connecting the two galaxies. Presently, there are no velocities of the stellar structure in this region, and so 
it is difficult to comment on any putative connection, although the numerical models suggest that these
features are unrelated \citep{2008MNRAS.390L..24B,2009Natur.461...66M}.

Figure~\ref{Figure3_M31} reveals the kinematic structure of the \ion{H}{1} in the outer disk and inner
halo of M31, where is an enhancement of \ion{H}{1} close to, but not
completely aligned where the GSS meets the disk of M31 (Section~\ref{halom31_space}). 
Intriguingly, while offset spatially, this 
gas is moving at $v_h$\avel{-510},  
similar to the velocity of the stellar content of the GSS, and with $v_{M31}$\avel{-200}.
At higher resolution, this gas is decomposed into distinct clumps; 
this will be discussed in Section~\ref{interpretation}.

 Also seen in stellar maps underlying Figures~\ref{Figure3_M31}  are Streams C and  D. 
 While there is apparently  no emission associated with Stream C, there is one blob
 of emission almost overlaying Stream D with $v_h$ \avel{-400}. This emission 
 arises close to two kinematic fields on Stream D obtained by \citet{2008MNRAS.390.1437C}, both
 of which yielded velocities of \vel{-391}; given the similarity of these velocities, and their distinct 
 difference from the large scale \ion{H}{1} gradient between M31 and M33 (with an expected velocity 
 of \avel{-250}) this is strongly suggestive of an association between the stellar and gaseous material 
 in Stream D.

\subsubsection{The disk \& halo of M33}\label{halom33_kine}
There is a gross alignment of the stellar debris and that of the \ion{H}{1} emission in the outskirts of M33, 
although there again is the spatial offset between the two components (Section~\ref{space} \& Figure~\ref{Figure4_M33}).
The kinematic properties of the gaseous material within the disk of M33 can be described with a galactic rotation model 
 \citep[i.e.][]{2003MNRAS.342..199C}, and also to the emission beyond edge of the stellar disk. 
Unfortunately, the only stellar kinematics obtained beyond the optical disk were obtained along the 
Southern major axis \citep{2006ApJ...647L..25M}, away from the prominent stellar debris  
\citep{2009Natur.461...66M,2010ApJ...723.1038M}, and hence no detailed comparison can be made.

\section{Interpretation}\label{interpretation}
What does the general lack of spatial and kinematic correlations between stellar substructure and \ion{H}{1} 
tell us about the  ongoing accretion in the M31-M33 system? 
Clearly, the axis connecting 
these two galaxies contains the majority of the substructure, but it appears that, in this case, the stellar and \ion{H}{1}
are the result of distinctly different accretion events; the main gaseous bridge through this region arises
from the interaction between M31 and M33 alone, with any stellar material too sparsely distributed 
to be detected \citep{2008MNRAS.390L..24B,2009Natur.461...66M}. Given M33's relatively large
mass, and potentially large impact-parameter, 
it has been able to retain the bulk of its stars and gas during the interaction with M31, although
the interaction has distorted both the stellar and gaseous disk of M33.

The progenitor of the Giant Stellar Stream, however would have been a lower mass system that has been interacting 
with M31, and, due to this interaction any progenitor gas was lost long ago, while
the remnant stellar debris exists in the form of tidal streams and shell-like caustics.   
At first glance, the lack of significant gaseous structures corresponding to the GSS and other associated stellar features may seem surprising.   However, as noted previously, while the gaseous and stellar components of an orbiting satellite are both subject to gravitational (tidal) forces, gas is also subject to shocks  and hydrodynamic drag.    
One possibility is that ram pressure stripping led to nearly complete loss gas from the GSS progenitor well before the 
formation of the observed GSS and associated structures.   
However, even if the ram pressure is not sufficiently efficient to completely strip the progenitor of its gas, 
hydrodynamic interactions can erase potential spatial and kinematic correlations between the two components.   
We are presently studying the gaseous and stellar streams  using high resolution numerical simulations (Shannon et al., in preparation) 
and preliminary results suggest that the two components can experience significant dislocation over the course of a single orbit, with the loss of angular momentum and energy due to hydrodynamic forces causing the gas to sink to the center of M31, unlike the much longer-lived stellar structures.

One intriguing correlation 
occurs where the GSS meets the disk of M31 (see Figure~\ref{Figure3_M31}). While 
the gas is spatially offset by \adist{15}, it velocity of $v_h$\avel{-500} is remarkably 
close to that of the GSS at this point. While they may be unrelated, 
it is worthwhile examining if it is plausible that they are physically conected. Such a question is 
complicated by the fact that there appears to be very little gas associated with the GSS, 
consistent with a lack of recent star-formation \citep{2006ApJ...652..323B},
but, this could be solved if the enhancement of \ion{H}{1} marks the (as yet unidentified)
progenitor of the GSS, with the offset in the spatial location illustrating the action of ram-pressure
stripping; such a conclusion would indicate that the accretion of the GSS is a relatively 
recent event and has not undergone several complete orbits, or by a progenitor 
massive enough to retain some of its gas through a prolonged accretion. This latter option 
is in agreement with the prediction of \citet{2008ApJ...682L..33F} who suggested that the source of the GSS 
was a rotating disk galaxy with a stellar mass of $\sim 10^9 {\rm M_\odot}$.

An alternate explanation also requires that the progenitor of the GSS was gas-rich, but that
some gas remains orbiting with the stars. 
This gas would be too tenuous to be visible in the observations presented in this paper, 
but as the stream is funnelled 
into the central regions, orbits converge and the local densities of both stars and gas increases.
Given this density enhancement, the gas could become visible to our observations. Again, the
offset in spatial location probably indicates ongoing ram-pressure, although there may be a delicate
balance between the amount of ongoing ram-pressure and the dynamical shepherding by the
associated stream of stripped dark matter, so that any can remain associated with the stream. Detailed
simulations of the formation and evolution of the GSS, considering realistic gas physics, are required to
address these questions.

Away from  M31, there is a tantalising correlation between \ion{H}{1} and stellar material 
in the vicinity of the SW Cloud. There are two potential interpretations for this, one with the 
SW Cloud representing an outwardly moving agglomeration of shredded stars, left over from a close interaction
with M31 $\sim250$ million years ago. This interaction would have to have been with a relatively pristine, 
gas-rich progenitor to leave the shredded debris with gas. On the other hand, the SW Cloud could represent the 
enhancement at the turning point of a stellar stream, and this stream would again 
have to be gas-rich to exhibit a similar enhancement in the \ion{H}{1}. 
Both scenarios require a relatively recent first passage of M31, ensuring that the progenitor 
does not lose all its gas through ram-pressure stripping through an extended interaction.
The estimated stellar and gas mass in the region of the SW Cloud are $\sim10^7 {\rm M_\odot}$ and
$\sim5\times10^5 {\rm M_\odot}$ respectively, suggesting that, if this is representative of a recent accretion
event, then the progenitor would have properties similar to the dwarf irregular IC10
\citep{alan2012a}.

While we do not have kinematics for the stellar component of the SW Cloud, it does appear to be spatially 
correlated with three globular clusters \citep[see][]{2010ApJ...717L..11M} for which velocities have been
recently determined (these will be presented in detail in forthcoming publications by \citet{mackey2012} 
and \citet{veljanoski2012}). If we take their velocities, of \avel{-433} (PAndAS-7), \avel{-411} (PAndAS-8) 
and \avel{-363} (PAndAS-14), each
with an error of less than \vel{10}, as being representative to that of the SW Cloud in general, we can compare
this gaseous spur velocity of \avel{-470}; which this is broadly consistent with the globular cluster velocity, especially if
we expect a velocity gradient along any putative stream connecting the two component. It should be noted, however, 
the width of the \ion{H}{1} in the SW Spur is \avel{200}, again indicative that we may be looking both outbound and 
inbound gases streams, although until we have stellar kinematics for the SW Cloud, and detailed dynamical modelling, 
the link between the SW Cloud and SW Spur remains circumstantial.

\begin{figure}
\begin{center}
\includegraphics[scale=1.0, angle=0]{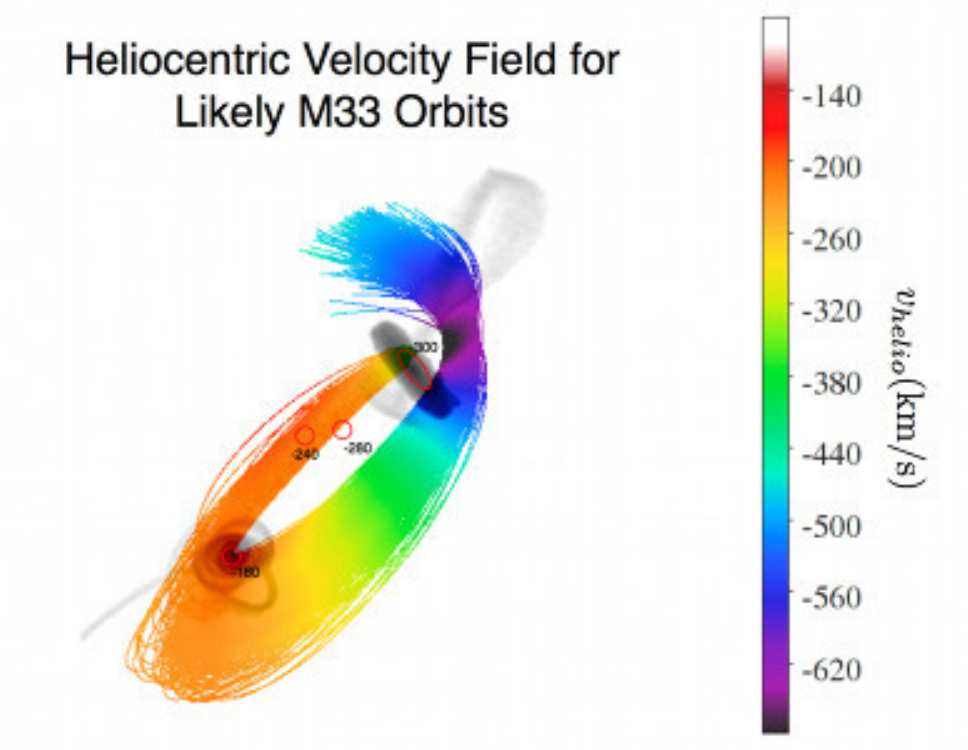}
\caption{The heliocentric velocity field from a distribution of likely M33 orbits
overlaid on an image from a simulation of a hypothetical interaction
between M33 and M31.  Also shown are the positions of some of the main gas
features from Fig.~\ref{Figure2_Whole_Map_HI} and their observed line of sight velocities.
The simulation contains only stars and dark matter without gas and shows
an enhanced view of the distorted outer disk of M33 and tidally stripped stars.
Some outer disk stars are thrown outward on a tidal tail to the SE while others
form a tidal bridge back towards M31.  The velocity field is generated from
a distribution of orbits generated from a Bayesian analysis that are
statistically consistent with the known distances, radial velocities
and proper motions of M31 and M33 and their given errors.  The orbits are
further constrained by the assumption that the M33 orbit passed within
30-60 kpc at pericentre assuming a mass model of M31 with $M=2.5\times
10^{12} {\rm M_\odot}$ and $r_{200}=280$ kpc with the plane of the orbit
passing close to the ``gas bridge'' between M31 and M33.  The close
agreement between the velocity of density peaks on the gas bridge and
orbital velocities of M33 in interacting models support the idea that the
bridge may be tidally stripped gas from M33 from a recent interaction.  A
caveat is that the pre-encounter gas disk of M33 would need to extend to
$\sim 15$ kpc.
}
\label{Figure6_M33}
\end{center}
\end{figure}

The distribution of gas within the M31-M33 system, especially the warped gas disk of M33 and the apparent gas bridge
connecting M31 to M33, strongly suggests a past interaction of
the two galaxies.   It has been argued that the
misalignment between the inner and outer gaseous disk of M33 and its
interpretation as a warped disk
could be the result of a recent tidal interaction (this is discussed in more detail below). The faint
extension of the stellar disk with a similar alignment as the outer gas disk can
be reproduced in interaction models \citep{2009Natur.461...66M}.  The gas bridge
between the two galaxies has also been modelled in the context of a past
interaction \citep[e.g.][]{2008MNRAS.390L..24B}.

While the interaction hypothesis is a plausible explanation for the phenomenology
of the gas in the M31-M33 system it ultimately depends on knowing the orbit of
M33.  Recent measurements of both the proper motion of M33 \citep{2005Sci...307.1440B}
and M31 \citep{2012ApJ...753....7S} provide strong constraints on the M33 orbit
within assumed mass models of M31.  We have been recently extending the
dynamical models described in \citet{2009Natur.461...66M} to describe the distorted
outer stellar disk of M33 to a range of mass models for M31 and M33 using a
Bayesian analysis to determine the distribution of orbits that are statistically
consistent with the observed distance, radial velocities and proper motions of
the galaxies.  This is a work in progress but we present some preliminary
results here relevant to the observed gas distribution.  In Figure \ref{Figure6_M33}, we
present the distribution of orbits and the resulting line-of-sight velocity
field consistent with the observations with the constraints that the pericentre
for the M33 orbit is in the range of \dist{30-60} in a mass model of M31 with
${\rm M}=2.5\times 10^{12} {\rm M}_\odot$ with $r_{200}=280$ kpc.  The orbits are
computed within the model potentials including a Chandrasekhar drag term to
model dynamical friction.  The drag coefficients are calibrated against live
N-body simulations to ensure their applicability to the Bayesian analysis.  We
further constrain the distribution to lie within 20 degrees of the plane
containing the gas bridge with the implicit assumption that the bridge is
created from tidally stripped gas from M33.  We have also overlaid the stellar
distribution from one N-body simulation including dark matter and stars without
gas to orient the orbital distribution as well as illustrate the expected
distortion of the outer disk of M33.  We note that the complex stellar features
seen around M33 in the simulation would be undetectable using current
observations but we present an enhanced view to illustrate the complex dynamics.
This model succeeds in two points in reference to the gas distribution.  First,
the model predicts that the outer disk of M33 will be warped and twisted as the
result of an interaction while the inner bright disk will remain unscathed.
Second, tidally stripped stars result in a tidal tail extending to the SE and a
tidal bridge that falls back onto M31.  While we have not modelled the gas in
this simulation, we expect stripped gas to follow approximately ballistic
trajectories similar to the stars.  The heliocentric velocity of the
distribution of orbits is consistent with the observed velocities of -240 and
-280 km/s seen in prominent gas features in the bridge.  A tidal bridge of gas
falling back onto M31 leading the position of M33 on its orbit after a past
interaction is a plausible description of these features.  

Finally, we look more closely at the stellar and gaseous material in the outskirts of M33, with
distinct features  visible in both the NW and SE (see Figure~\ref{Figure4_M33}).
The position angle of the \ion{H}{1} feature is offset from that of the disk by $\sim 30^\circ-40^\circ$, 
and has been interpreted as
a warped disk in projection and modelled in terms of tilted rings by \citet{1997ApJ...479..244C}.
The stellar features  lie roughly between the disk and \ion{H}{1} position angles.

The origin of both the \ion{H}{1} and stellar features has been attributed to tidal interactions with M31 during a close encounter 
(\citealt{2009ApJ...703.1486P}; \citealt{2009Natur.461...66M}, but see \citealt{Reakes1978} for an early discussion 
of this possibility), though models of a steady state precessing warp in a model
with a flattened halo have also been considered \citep[e.g.][]{1991ApJ...376..467K}.  
Competing with tidal effects is the gravitational field of the M33 disk and dark
halo.  In addition, the \ion{H}{1} disk experiences ram pressure due to its interaction with the gaseous halo of M31. 
We can distinguish between three regions: The outermost part of the disk where tidal interactions and ram pressure strip material from M33 leaving behind a stream, which roughly traces the M33 orbit; an intermediate region, where material is stripped from the M33 disk but remains bound to the galaxy; and a close-in region, where the disk is warped but material remains on roughly circular orbits.

\begin{figure}
\begin{center}
\includegraphics[scale=1.0, angle=0]{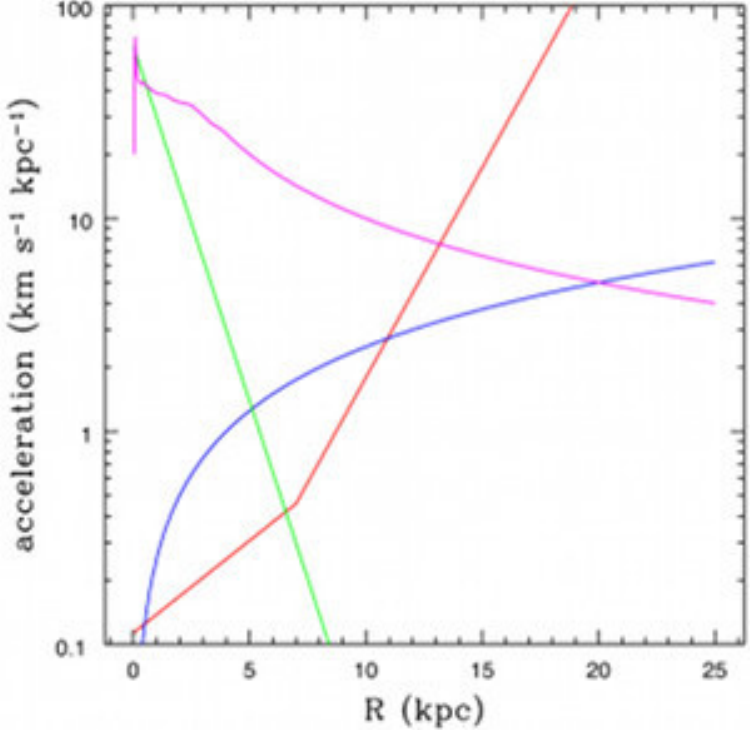}
\caption{The various components of  force (per unit mass) for the restoring force of M33 (magenta) and the disk (green),
ram pressure (red) and the M31 tidal field (blue).  At the tidal radius, $R_t \simeq 20 \,{\rm kpc}$, the forces due to the M31 
and M33 tidal fields are equal, while ram-pressure stripping can disrupt the gaseous disk at $R\gtrsim 7\,{\rm kpc}$ and
completely strip material from beyond $R\gtrsim 14\,{\rm kpc}$.}
\label{Figure_7}
\end{center}
\end{figure}

In this simplified argument, we consider an element of the M33 gas disk at galactocentric radius $R$ 
with surface area $\Delta \Omega$ and surface density $\Sigma_{\rm gas}$.  The tidal force on this element due to M31 is 
\begin{equation}
F_{\rm tid} \simeq \Sigma_{\rm gas}
\frac{V_{M31}^2}{R_p^2}R \Delta \Omega
\end{equation}
where $R_p$ is the perigalactic distance between M31 and M33.  In deriving this expression we have assumed that $R_p\gg R$ and that M31 has a flat rotation curve with a circular velocity $V_{M31}$.
The ram pressure force is
\begin{equation}
F_{rp} \simeq \rho_{\rm gas} V_{\rm rel}^2 \Delta \Omega
\end{equation}
where $\rho_{\rm gas}$ is the density of the M31 gaseous halo.
The gravitational restoring force due to the M33 disk is
\begin{equation}
F_{d,{\rm res}} \simeq 4\pi G\Sigma_{\rm disk} \Sigma_{\rm gas}\Delta \Omega
\end{equation}
where $\Sigma_{\rm disk}$ is the total surface density of the disk.  If we assume an 
exponential disk, then $\Sigma_{\rm disk} = \left (M_d/4\pi G R_d\right )\exp\left (-R/R_d\right )$.  The 
restoring force due to M33 as a whole is 
\begin{equation}
F_{t,{\rm res}} = \frac{V_{M33}^2}{R}\Sigma_{\rm gas} \Delta \Omega
\end{equation}
where we assume that M33 also has a flat rotation curve.  
Setting $F_{\rm tid} = F_{t,{\rm res}}$ yields an estimate for the tidal radius: $R_t \simeq R_p\left (V_{M33}/V_{M31}\right )$.

The different forces are shown in Figure~\ref{Figure_7}, and for purely illustrative purposes, we set $V_{M31} = 250 {\rm km\,s}^{-1}$, $V_{M33} = 100 {\rm km\,s}^{-1}$, $V_{\rm rel} = 400\,{\rm km\,s}^{-1}$, $R_p = 50\,{\rm kpc}$, $M_d = 2.6\times 10^9\,{\rm M_\odot}$, $R_d = 1.1\,{\rm kpc}$, and $\rho_{\rm gas} = 8\times 10^{-5}\,{\rm cm}^{-3}m_H$.  The forces due to the M31 tidal field and due to M33 itself are equal at the tidal radius $R_t \simeq 20 \,{\rm kpc}$.  Ram pressure can disrupt the disk at $R\gtrsim 7\,{\rm kpc}$ and strip material from M33 $R\gtrsim 14\,{\rm kpc}$.

While these arguments  represent order-of-magnitude estimates, they appear to roughly correspond to the 
observed structures in M33. However, a full exploration of the M31-M33 interaction, especially with regards to 
the stripping of M33, using models including gas, stars and dark matter, are required to fully understand the 
complex dynamics on display; this will form the basis of an upcoming paper (Dubinski et al. 2012). 

\section{Conclusions}\label{conclusions}
We have presented the spatial and kinematic correlations of the stellar and gaseous 
substructure within the haloes of M31 and M33. While significant
substructure is apparent in each component, and there is a gross alignment of significant features,
mainly along the axis connecting M31 and M33,  there is
a distinct lack of correlation between the detailed structures. 

The lack of a truly global kinematic survey of the stellar substructure within the halos of M31 and M33 
limits a detailed comparison of stars and gaseous material, although the growing number of
``key-hole'' observations  with 10-m class telescopes is addressing this.
However, this situation will be resolved with 
the advent of wide-field spectroscopy on large telescopes, such as the proposal to build the 
ngCFHT\footnote{\tt orca.phys.uvic.ca/$\sim$pcote/ngcfht}, 
ushering in a new era in understanding galactic archaeology.

The resulting conclusion of this study, therefore, is that the gaseous and stellar substructure within
the halos of M31 and M33 have been built through a number of distinct accretion events, but 
differing physical processes have driven
the disruption of the stellar and gaseous components of any particular substructure, 
with the latter suffering the additional
forces of shocking and ram-pressure stripping. Given that, through the use of high-resolution 
numerical simulations, our understanding of the differing processes driving the disruption of stellar and 
gaseous material, a comparison of their distribution through the halo should allow us 
to dynamically date the accretion; this will be the subject of a 
future contribution.

\acknowledgments
GFL gratefully acknowledges financial support for his ARC Future Fellowship
(FT100100268) and through the award of an ARC Discovery Project (DP110100678). 
Furthermore, GFL thanks the Institute of Astronomy at the University of Cambridge
for their support and hospitality during his sabbatical during the latter half of
2010, where a substantial portion of this work was undertaken.

\end{document}